\documentstyle[fleqn]{article}
\begin{document}
\newcommand \s[1] {\mbox{\hspace{#1cm}}}
\newcommand \FR[2]
 {\s{.06}{\strut\displaystyle#1\over\strut\displaystyle#2}\s{.06}}
\newcommand \fr[2] {\s{.06}{\textstyle{#1\over#2}}\s{.06}}
\newcommand \eqnum[1]{\eqno (#1)}
\renewcommand \d {{\rm d}}
\hyphenation{counter-term counter-terms}

\begin{center}

{\LARGE \bf Noninvariant renormalization \medskip \protect \\
 in the background-field method\footnote{Supported in part by RFFR grant
 \# 94-02-03665-a}}

\bigskip

{\large L. V. Avdeev\footnote{E-mail: $avdeevL@thsun1.jinr.dubna.su$},
 D. I. Kazakov\footnote{E-mail: $kazakovD@thsun1.jinr.dubna.su$}} 
and~
$\underline{\mbox{\large M. Yu. Kalmykov\footnote{E-mail:
 $kalmykov@thsun1.jinr.dubna.su$}}}$

\bigskip

{\em Bogoliubov Laboratory of Theoretical Physics,\\
 Joint Institute for Nuclear Research,\\
 $141\,980$ Dubna $($Moscow Region$)$, Russian Federation}

\end{center}

\begin{abstract}

We investigate the consistency of the back\-ground-field formalism when
applying various regularizations and renormalization schemes. By an
example of a two-di\-men\-sion\-al $\sigma$ model it is demonstrated
that the back\-ground-field method gives incorrect results when the
regularization (and/or renormalization) is noninvariant. In particular,
it is found that the cut-off regularization and the differential
renormalization belong to this class and are incompatible with the
back\-ground-field method in theories with nonlinear symmetries.

\end{abstract}

\section{Introduction} \indent

Using a renormalization procedure in gauge theories, one has to be very
careful not to violate the gauge invariance on the quantum level, thus
destroying the renormalizability of the model. Therefore, one is bound
to apply an {\em invariant} renormalization. By this we mean a
renormalization that preserves all the relevant symmetries of the model
on the quantum level, that is, preserves all the Ward identities
\cite{Sl} for the renormalized Green functions.

On the one hand, this can be achieved by applying an invariant {\em
regularization} first \cite{invariant} (respecting the symmetries in the
regularized theory) and then using, for instance, the minimal
subtraction scheme \cite{min,Sp} to fix the finite arbitrariness in
local counterterms. On the other hand, when the regularization is
noninvariant or no explicit regularization is introduced at all, there
is no {\em automatic} preservation of the symmetries. Then one has to
take care of that directly. For example, one can use some noninvariant
regularization and consecutively choose certain finite counterterms to
restore the invariance (the symmetry of the renormalized Green
functions) order by order in perturbation theory~\cite{B}.

When treating theories with nonlinear realization of a symmetry, like
two-di\-men\-sion\-al $\sigma$ models or quantum gravity, one faces
extraordinary complexity of perturbative calculations. To simplify them,
one usually applies the so-called back\-ground-field method \cite{back}
which allows one to handle all the calculations in a strictly covariant
way. This method was successfully applied to multiloop calculations in
various gauge and scalar models, being combined with the minimal
subtraction scheme based on some invariant regularization.

The most popular and handy regularization used in these calculations was
the dimensional regularization \cite{dim}. It has been proved to be an
invariant regularization, preserving all the symmetries of the classical
action that do not depend explicitly on the space-time dimension
\cite{Sp,action}. Moreover, any formal manipulations with the
dimensionally regularized integrals are allowed. However, an obvious
drawback of this regularization is the violation of the axial invariance
and of supersymmetry. That is why there are numerous attempts to find
some other regularization equally convenient and efficient. Among such
schemes the recently proposed differential renormalization \cite{diff}
is discussed.

In the present paper we investigate the compatibility of the
back\-ground-field formalism with various regularizations and
renormalization prescriptions. Our conclusion is that the
back\-ground-field method necessarily requires one to use an invariant
renormalization procedure. As the invariance does not hold, the method
gives incorrect results. We demonstrate this by an example of the
two-di\-men\-sion\-al $O(n)$ $\sigma$ model, comparing the dimensional
regularization, the cut-off regularization within the minimal
subtraction scheme, and the differential renormalization method.

\section{Two-dimensional nonlinear $O(n)~ \sigma $ model} \indent

Let us consider the two-di\-men\-sion\-al $\sigma$ model of the $O(n)$
principal chiral field (${\bf n}$ field) and calculate the two-loop
$\beta$ function, using various approaches. The model is described by
the lagrangian
\begin{equation} \label{n-field} {\cal L} = \fr 1 2 h^{-1}
 \left( \partial_\mu {\bf n} \right) ^2, \s2 {\bf n}^2 = 1.
\end{equation}
It can be treated as a special case of the generic bosonic $\sigma$
model
\begin{equation} \label{sigma}
 {\cal L} = \fr 1 2 (\partial_\mu \phi^j)~ g_{jk}(\phi)~
 \partial_\mu \phi^k , \s2 (j,k=1,2,...,n-1) ,
\end{equation}
where the metric is of the form
\begin{equation} \label{metric}
 g_{jk}(\phi)= \delta_{jk} +h~ \phi_j\phi_k / (1-h~\phi^2) .
\end{equation}

The back\-ground-field expansion of the action can be done in a strictly
covariant fashion~\cite{sigma}. To separate the ultraviolet and infrared
divergencies, we add an auxiliary mass term to eq.~(\ref{sigma})
\begin{equation} \label{mass}
 {\cal L}_m= \fr 1 2 m^2~ \phi^j~ g_{jk}(\phi)~ \phi^k .
\end{equation}
This additional term serves only for eliminating infrared divergencies,
naively present in any two-di\-men\-sion\-al theory with massless
scalars. After calculating ultraviolet logarithms, one should set
$m^2$=0.

\begin{figure}[htp] % 1

 $$ \FR 1 2 R_{jk}~ (\partial_\mu \phi^j)~ (\partial_\mu \phi^k) \s{.4}
  \begin{picture}(10,7)(-5,-2.5) % simple tadpole
   \put(0,0){\circle 7}
   \put(0,-3.5){\line(-1,-1){3.6}}
   \put(0,-3.5){\line(1,-1){3.6}}
   \put(0,-3.5){\circle*1}
  \end{picture} \eqnum{\rm a} $$
 $$ - \FR 1 {12}
  \left( 2~ {R^a}_j~ R_{ak} + 3~ {R^{abc}}_j~ R_{abck}
  \right) (\partial_\mu \phi^j)~ (\partial_\mu \phi^k) \s{.4}
  \begin{picture}(10,7)(-5,-1) % double tadpole
   \put(0,2.8){\circle{5.6}}
   \put(0,-2.8){\circle{5.6}}
   \put(-4,0){\line(1,0)8}
   \put(0,0){\circle*1}
  \end{picture} \eqnum{\rm b} $$
 $$ + \FR 1 6 {R_j}^{ab}{}_k~ R_{ab}~
  (\partial_\mu \phi^j)~ (\partial_\mu \phi^k) \s{.4}
  \begin{picture}(10,8)(-5,-1) % frog
   \put(0,0){\circle{5.6}}
   \put(0,5.6){\circle{5.6}}
   \put(-2.8,0){\vector(0,-1){.7}} % down
   \put(2.8,0){\vector(0,-1){.7}}
   \put(0,0){\circle*{.6}}
   \put(0,2.8){\circle*1} % vertices
   \put(0,-2.8){\circle*1}
   \put(0,-2.8){\line(-1,-1){3.6}} % legs
   \put(0,-2.8){\line(1,-1){3.6}}
  \end{picture} \eqnum{\rm c} $$
 $$ - \FR 4 9 {R_j}^{(ab)c}~ R_{kabc}~
  (\partial_\mu \phi^j)~ (\partial_\nu \phi^k) \s{.4}
  \begin{picture}(16,9)(-8,-1) % 2-loop nut
   \put(0,0){\circle{9.8}}
   \put(-7,0){\line(1,0){14}}
   \put(-4.9,0){\circle*1}
   \put(4.9,0){\circle*1}
   \put(-2.5,0){\vector(1,0){.7}} % one line with two arrows
   \put(2.5,0){\vector(-1,0){.7}}
  \end{picture} \eqnum\d $$
 $$ - \FR 8 9 {R_j}^{(ab)c}~ R_{k(cb)a}~
  (\partial_\mu \phi^j)~ (\partial_\nu \phi^k) \s{.4}
  \begin{picture}(16,8)(-8,-1) % 2-loop nut
   \put(0,0){\circle{9.8}}
   \put(-7,0){\line(1,0){14}}
   \put(-4.9,0){\circle*1}
   \put(4.9,0){\circle*1}
   \put(-3.5,3.5){\vector(1,1){.5}} % two lines with an arrow
   \put(3.5,-3.5){\vector(-1,-1){.5}}
  \end{picture} \eqnum{\rm e} $$

 \caption{The one- and two-loop corrections to the effective action of
  the two-di\-men\-sion\-al bosonic $\sigma$ model without torsion.
  Lines of the diagrams refer to propagators $1/(p^2$+$m^2)$, and arrows
  to $p_\mu$ in numerators.}

\end{figure}

The $\sigma$ model (\ref{sigma}) with the particular choice of the
metric (\ref{metric}) becomes renormalizable. All the covariant
structures that may appear as counterterms are reducible to the metric,
so that the only thing that happens is a renormalization of the kinetic
term. By rescaling the fields the renormalization can be absorbed into
the charge. The invariant charge $\overline{h}$=$Z^{-1}h$ is defined
through the field renormalization constant $Z$. To calculate $Z$ within
the back\-ground-field method, one has to consider the
one-particle-irreducible diagrams with two external lines of the
background field, and quantum fields inside the loops. Up to two loops
the relevant diagrams \cite{Bel} are shown in fig.~1. Their
contributions to $Z$ are obtained by normalizing to the tree term
$(-\fr 1 2)~ (\partial_\mu \phi^j)$ $g_{jk}$ $\partial_\mu \phi^k$.

The Riemann and Ricci tensors are the functionals of the background
field. In our model with the metric given by eq.~(\ref{metric}) they are
evaluated to
\begin{equation} R_{abjk}= h \left( g_{aj}~g_{bk} -g_{ak}~g_{bj} \right)
 , \s2 R_{jk}=(n-2)~h~g_{jk}~. \nonumber
\end{equation}

Besides, we should take into account the renormalization of the mass
$Z_{m^2}$. In the first loop it is determined by the diagram of fig.~2
[normalized to $(-\fr 1 2)~ m^2~ \phi^j~ g_{jk}~ \phi^k$]. Although
operator (\ref{mass}) gives no direct contribution to the
wave-func\-tion renormalization, in all the diagrams that contribute to
$Z$ the mass ought to be shifted by such corrections. In the two-loop
approximation, only the fig.-2 correction to fig.~1 is essential.

\begin{figure}[htp] % 2
 $$ \FR 1 6 m^2~ R_{jk}~ \phi^j \phi^k \s{.4}
  \begin{picture}(10,7)(-5,-2.5) % simple mass tadpole
   \put(0,0){\circle 7}
   \put(0,-3.5){\line(-1,-1){3.6}}
   \put(0,-3.5){\line(1,-1){3.6}}
   \put(0,-3.5){\circle*1}
  \end{picture} $$
 \caption{The one-loop mass correction to the effective action.}
\end{figure}

One can find the $\beta$ function by requiring independence of the
invariant charge on the normalization point. This leads to the following
expression for the $\beta$ function through the finite wave-function
renormalization constant:
\begin{equation} \label{beta} \beta (h)= h
 \left( \mu^2 \FR {\partial Z} {\partial \mu^2}
 \right)
 \left[ \Bigl( 1-h \FR \partial {\partial h} \Bigr) Z
 \right] ^{-1} .
\end{equation}

Using the dimensional regularization and the minimal subtraction scheme
to calculate the diagrams presented above, one obtains the following
well-known expression for the two-loop $\beta$ function of the ${\bf
n}$-field model (\ref{n-field}) \cite{n-field,Bel}:
\begin{equation} \label{beta_dim}
 \beta_{\rm dim}(h)= -(n-2) \Big[ 1 +2~ h/(4\pi) \Big] h^2/(4\pi) .
\end{equation}

The dimensional regularization and the minimal subtraction scheme 
provide us with an invariant renormalization procedure \cite{Sp,action}
within the back\-ground-field method. Hence, the obtained expression for
the $\beta$ function is correct, and we can use it as a reference
expression to compare with other approaches. Owing to the presence of
just one coupling constant in the model, the $\beta$ function should be
renormalization-scheme independent up to two loops and should coincide
with eq.~(\ref{beta_dim}).

Let us check the validity of the back\-ground-field method in
conjunction with other regularizations and renormalization
prescriptions.

\subsection{The Cut-Off Regularization} \indent

We start with the regularization that uses a cut-off in the momentum
space. All the integrals over the radial variable in the Euclidean space
are cut at an upper limit~$\Lambda$. Strictly speaking, this is not a
very promising regularization, since it explicitly breaks the Lorentz,
as well as gauge, invariance. However, we use it here to realize what
may happen when a noninvariant regularization is applied.

When the regularization parameter has the dimension of a mass, the
minimal subtraction procedure can be defined \cite{log} so as just to
convert the logarithms of the (infinite) cut-off~$\Lambda$ into the
logarithms of a finite renormalization point $\mu$ which appears in the
theory after renormalizations:
\begin{equation} \label{K}
 {\cal K}~ \ln^n (\Lambda^2)= \ln^n (\Lambda^2) -\ln^n (\mu^2),
 \s2 {\cal K}~ \Lambda^n = \Lambda^n ,
\end{equation}
so that
\begin{equation} \label{R}
 {\cal R}~ \ln^n (\Lambda^2) = (1-{\cal K})~ \ln^n (\Lambda^2) =
 \ln^n (\mu^2), \s2 {\cal R}~ \Lambda^n = 0.
\end{equation}
In case of overlapping divergencies, which generate powers of the
logarithms, one ought to perform the standard renormalization procedure
prior to the subtractions. However, any contributions of the minimally
subtracted counterterms (\ref{K}) will be annihilated by (1$-$${\cal
K}$), eq.~(\ref{R}), irrespective of any powers of the logarithms from
the residual graphs with contracted subgraphs.

Thus, it is sufficient to calculate the regularized diagrams of figs.~1
and 2, up to $\Lambda$-power corrections and ultra\-violet-finite
two-loop contributions, and then to replace $\Lambda^2 $ with $\mu^2 $,
and $m^2$ in the one-loop diagram of fig.~1a with $m^2 Z_{m^2}$,
including the correction from fig.~2. The charge-renormalization
constant proves then to be
\begin{equation} \label{Z_cut}
 Z_{\rm cut}= 1 -(n-2)~ h/(4\pi)~ \ln (\mu^2/m^2) +0\cdot h^2,
\end{equation}
so that eq.~(\ref{beta}) gives the $\beta$ function
\begin{equation} \label{beta_cut}
 \beta_{\rm cut}(h)= -(n-2)~ \Big( 1+ 0\cdot h \Big)~ h^2/(4\pi).
\end{equation}

The difference between this result and that obtained in dimensional
renormalization (\ref{beta_dim}) is a direct manifestation of the
noninvariance of the cut-off regularization, which violates the
translational invariance. However, one needs to explain the reason for
the failure to reproduce the correct $\beta$ function in the present
case. Although the cut-off regularization is noninvariant, still it has
been successfully used to perform multiloop calculations in scalar field
theories and in the quantum electrodynamics both within the
back\-ground-field method and by the conventional diagram technique.

The point is that those theories were renormalizable in the ordinary
sense, that is, they had a finite number of types of divergent diagrams.
In contrast, the ${\bf n}$-field model is renormalizable only in the
generalized sense. The total number of divergent structures here (with
various external lines) is infinite, but they are related to each other
by general covariance of the renormalized theory (in case of an
invariant renormalization). So the number of independent structures
remains finite. Expanding the lagrangian, we get an infinite number of
terms; however, the renormalization constants are not arbitrary but
mutually related. Although the back\-ground-field method formally
preserves the covariance of the model, the use of a noninvariant
renormalization would break the intrinsic connection between various
diagrams (and between their renormalization constants), thus leading to
wrong results.

Therefore, we conclude that in generalized renormalizable (as well as
nonrenormalizable) theories it is not allowed to use the cut-off
regularization with the minimal subtractions in the framework of the
back\-ground-field method.

\subsection{Differential Renormalization} \indent

We now want in the same way to check the invariance properties of the
differential renormalization method. Its idea traces back to the
foundations of the renormalization procedure \cite{Bog} as a
re-defi\-ni\-tion of the product of distributions at a singular point.
The method suggests to work in the co-ordinate space, where the free
Green functions are well defined, although their product at coinciding
points suffers from ultraviolet divergencies. The divergencies manifest
themselves as singular functions which have no well-defined Fourier
transform. The recipe of the differential renormalization \cite{diff}
consists in rewriting a singular product in the form of a differential
operator applied to a nonsingular expression:
\begin{equation} \label{rewrite}
 f(x_j,...,x_k)= D
 \big( \Box^{\sigma_j}_{x_j},...,\Box^{\sigma_k}_{x_k} \big)~
 g(x_j,...,x_k) ,
\end{equation}
Eq.~(\ref{rewrite}) should be understood in the sense of distributions,
that is, in the sense of integration with a test function. Then one
ignores any surface terms on rearranging the derivatives via integration
by parts. The nonsingular function $g(x_j,...,x_k)$ is obtained by
solving a differential equation, and hence, involves an obvious
arbitrariness. The latter can be identified with the choice of a
renormalization point and a renormalization scheme. In this respect the
differential renormalization does not differ from any other
renormalization prescription.

In the absence of a primary regularization this prescription might
preserve all the needed invariances and, what is important for
applications, seems to renormalize ultraviolet singularities in the
integer dimension. On the other hand, the absence of any intermediate
regularization prevents one from using the standard scheme: invariant
regularization + minimal subtractions. Therefore, to verify the
invariance properties of the differential renormalization, one has to
deal with renormalized amplitudes directly.

Two-loop calculations of the renormalization constant in the
two-di\-men\-sion\-al $\sigma$ model in the framework of the
differential renormalization have been performed in
ref.~\cite{troubles}. The authors have used the concept of the infrared
$\widetilde{\cal R}$ operation to handle the infrared divergencies. 

An important role in the calculations plays the tadpole diagram
(fig.~1a). In four dimensions, diagrams of this type diverge
quadratically and can be consistently renormalized to zero, as it was
originally done in the method of the differential renormalization
\cite{diff}. However, in two dimensions the leading one-loop
contribution to the $\beta$ function comes from this very diagram.
Hence, the tadpole should be different from zero in any renormalization.
This means that we have to define the two-di\-men\-sion\-al tadpole
diagram in a self-con\-sist\-ent way {\em in addition} to the recipe of
the differential renormalization. Such an extension, based on the
assumption of invariance, has been discussed in detail in
ref.~\cite{troubles} for the massless case. The expression for the
$\beta$ function has been found to be
\begin{equation} \label{beta_diff0}
 \beta_{\rm diff}(h)= -(n-2)~ h^2/(4\pi) +0\cdot h^3 .
\end{equation}
Thus, the suggested definition of the massless tadpole leads to the
correct expression for the {\em one}-loop $\beta$ function, but fails in
two loops.

Therefore, we would like to circumvent possible ambiguities of combining
the infrared $\widetilde{\cal R}$ operation with the differential
renormalization. We are going to apply the method to the massive model
(\ref{mass}) in which no infrared difficulties ever appear. The free
propagator is of the form
\begin{equation} \label{D_m} \Delta_m (x)= (2\pi)^{-1} K_0(m |x|),
\end{equation}
where $K_0$ is the MacDonald function. Bearing in mind the known
expansion
\begin{equation}
 K_0(x)= -\ln \Bigl( \FR x 2 \Bigr) \sum_{k=0}^{\infty}
 \FR 1 {(k!)^2} \Bigl( \FR x 2 \Bigr) ^{2k}
 + \sum_{k=0}^{\infty} \FR {\psi(k+1)} {(k!)^2}
 \Bigl( \FR x 2 \Bigr) ^{2k} , \nonumber
\end{equation}
by analogy with ref.~\cite{troubles} we infer the following definition
for the massive tadpole:
\begin{equation} \label{tad1}
 {\cal R} \Big[ \delta(x)~ \Delta_m (x) \Big] = \delta(x) /(4\pi)~
 \ln(M^2/m^2) .
\end{equation}

In the spirit of the consistent ${\cal R}$ operation, the squared
tadpole (fig.~1b) should be defined as the square of the renormalized
value (\ref{tad1}) for fig.~1a, that is,
\begin{equation} \label{tad2}
 {\cal R} \Big[ \delta(x)~ \Delta_m^2 (x) \Big] = \delta(x)
 \left[ (4\pi)^{-1} \ln(M^2/m^2) \right] ^2 .
\end{equation}

Below we present the contributions of all the diagrams to the
renormalization constants:
\begin{eqnarray*}
Z({\rm fig.~1a}) &=& -(n-2)~ h/(4\pi)~ \ln(M^2/m^2), \\
Z({\rm fig.~1b}) &=& \fr 1 3 (n-2)~(n+1)~ h^2/(4\pi)^2~
 \ln^2 (M^2/m^2), \\
Z({\rm fig.~1c}) &=& -\fr 1 3 (n-2)^2~ h^2/(4\pi)^2~ \ln(M^2/m^2)
 \left[ \ln (M_1^2/m^2) -1 \right] , \\
Z({\rm fig.~1d}) &=& -\fr 2 3 (n-2)~ h^2/(4\pi)^2~ \ln^2 (M_2^2/m^2), \\
Z({\rm fig.~1e}) &=& -\fr 1 3 (n-2)~ h^2/(4\pi)^2~ \ln^2 (M_3^2/m^2), \\
Z_{m^2}({\rm fig.~2}) &=& -\fr 1 3 (n-2)~ h/(4\pi)~ \ln(M^2/m^2),
\end{eqnarray*}
where we introduce different ultraviolet renormalization scales for 
different diagrams~\cite{all}. Thus,
\begin{equation} \label{beta_diff}
 \beta_{\rm diff}(h)= -(n-2) \FR {h^2} {4\pi}
 \left\{ 1 +\FR h {4\pi}
  \left[ \FR 1 3 (n-2)~ \ln \FR {M_1^2} {M^2}
   +\FR 4 3 \ln \FR {M_2^2} {M^2} +\FR 2 3 \ln \FR {M_3^2} {M^2}
  \right]
 \right\} .
\end{equation}
We see that the result explicitly depends on the ratio of the
renormalization scale parameters in different diagrams. Such a
dependence on the details of the renormalization prescription is beyond
the usual scheme arbitrariness. It would never occur to two loops in the
conventional perturbation theory for ordinary renormalizable one-charge
models. There the arbitrariness would be completely absorbed into a
finite number of counterterms which are of the operator types present in
the tree lagrangian. Hence, we should try to fix the parameters of the
differential renormalization by imposing some additional requirements.
In the quantum electrodynamics the gauge Ward identities could be used
to this end \cite{diff}. For the $\sigma$ model in the
back\-ground-field formalism the situation is not so clear.

The parameter $M_1$ that appears in the one-loop tadpole subgraph of
fig.~1c with the numerator can be fixed as follows. In the momentum
representation we can easily see that the sum of this diagram and the
simple tadpole (fig.~1a) is just an ultra\-violet-finite integral which
equals $1/(4\pi)$. The value will be correctly reproduced by the
differential renormalization if we choose the same scale for both
tadpole graphs: $M_1$=$M$. Thus, for these diagrams the renormalization
seems to be automatically invariant.

Assuming the complete automatic invariance, we would set $M_2$=$M_3$=$M$
as well. However, then eq.~(\ref{beta_diff}) would again give us the
wrong result (\ref{beta_diff0}) obtained under the same assumption in
the massless theory via the infrared $\widetilde{\cal R}$ operation.
Hence, the ratio of the renormalization parameters ought to be somehow
tuned in order to restore the invariance.

The identical situation was encountered in a nonrenormalizable chiral
theory already at the one-loop level for physical observables
\cite{chiral}. Inside the differential renormalization, one finds no
{\em a~priori} internal criterion for choosing the ratios of auxiliary
masses, to get reliable results. Of course, comparing
eq.~(\ref{beta_diff}) to eq.~(\ref{beta_dim}) in the dimensional
renormalization [or the results for fig.~1(d,e) individually], we can
infer the values that would ensure the invariance:
$\ln(M_2^2/M^2)=\ln(M_3^2/M^2)=1$. But by itself the differential
renormalization remains ambiguous, if we apply it to a theory that is
not renormalizable in the ordinary sense. Thus, it is not directly
compatible with the back\-ground-field method.

\section{Conclusion} \indent

Our examples show that the back\-ground-field formalism requires one to
use an invariant renormalization procedure in order to obtain valid
results in a gen\-er\-al\-ized-re\-norm\-al\-iz\-able theory. A
noninvariant regularization or renormalization may break an implicit
correlation between different diagrams, which is essential as one
formally expands the action in the background and quantum fields.

We have demonstrated by direct two-loop calculations that the
regularization via a cut-off in the momentum space is noninvariant and
gives a wrong result for the $\beta$ function of the ${\bf n}$-field
model within the back\-ground-field formalism.

We have also found that the differential renormalization is not
automatically invariant. The result depends on the ratio of the
auxiliary scale parameters beyond the allowed scheme arbitrariness in
the second order of perturbation theory. We can partially fix the
ambiguity by imposing a condition on divergent one-loop tad\-pole-type
diagrams a combination of which should be finite. But this is not
enough, and there seems to be no algorithm of generalizing such
conditions to more complicated graphs.

We would like to stress once more that the calculations in nonlinear
models like the $\sigma $ model or supergravity are hardly possible
without the background-field formalism. Thus, the need in the
regularization that preserves the underlying symmetries and is
practically convenient at the same time is of vital importance. The
above examples clearly demonstrate the difficulties that arise when one
uses a noninvariant procedure.

\end{document}